\documentclass[conference,a4paper]{article}

\usepackage[margin=1.2in]{geometry}
\usepackage[utf8]{inputenc} 
\usepackage[T1]{fontenc}
\usepackage{url}
\usepackage{ifthen}
\usepackage{cite}
\usepackage{todonotes}                  
\usepackage{color,soul}    
\usepackage[cmex10]{amsmath} 

\usepackage{./rmylay,flushend}

\newcommand{\x}{\textbf{x}}
\newcommand{\y}{\textbf{y}}
\newcommand{\w}{\textbf{w}}
\newcommand{\s}{\textbf{s}}
\newcommand{\M}{\mathcal{M}}
\newcommand{\iid}{\emph{i.i.d.}\xspace}
\newcommand{\hatp}{\hat{P}}

\begin{document}
\title{Redundancy of Markov Family with Unbounded Memory} 

\author{%
  Changlong Wu\qquad Maryam Hosseini\qquad Narayana Santhanam\\
  University of Hawaii at Manoa\\
                    Honolulu\\ 
                    96822 Hawaii, USA\\
                    Email: \{wuchangl, hosseini, nsanthan\}@hawaii.edu
}
\maketitle

\maketitle

\begin{abstract}
 We study the
  redundancy of universally compressing strings $X_1\upto X_n$
  generated by a binary Markov source $p$ without any bound on the
  memory. To better understand the connection between
  compression and estimation in the Markov regime, we consider a class
  of Markov sources restricted by a \emph{continuity} condition. In
  the absence of an upper bound on memory, the continuity condition
  implies that $p(X_0|X^{-1}_{-m})$ gets closer to the true
  probability $p(X_0|X_{-\infty}^{-1})$ as $m$ increases, rather than
  vary around arbitrarily.  For such sources, we prove asymptotically
  matching upper and lower bounds on the redundancy. In the process,
  we identify what sources in the class matter the most from a
  redundancy perspective.
\end{abstract}

\section{Introduction}
Estimation and compression are virtually synonymous in the \iid
regime.  Indeed, in the \iid case, the add-half (and other
add-constant) estimators that provide reasonable estimates of
probabilities of various symbols are described naturally using
a universal compression setup. These estimators simply correspond to
the conditional probabilities assigned by a Bayesian mixture when
Dirichlet priors on the parameter space---and indeed encoding the
probabilities given by these Bayesian mixtures yields good
universal compression schemes for these classes of distributions.

In the Markov setup, there is an additional complication not seen
in the \iid setup---\emph{mixing}---that complicates the relation between
compression and estimation. Without going into the technicalities
of mixing, slow mixing sources do not explore the state space efficiently.

For example, consider a memory-1 binary Markov source that assigns
conditional probability of $1-\epsilon$ for a 1 given 1, and
$\epsilon$ for the conditional probability of a 1 given 0. If we start
the source from the state $1$, for sample length $n\ll\frac1\epsilon$,
we will see a sequence of all 1s with high probability in length-$n$
samples. This sequence of 1s is, of course, easy to compress---but
clearly precludes the estimation of the conditional probabilities
associated with 0. 

Previous work in the Markov regime, however, has typically considered
classes Markov sources with bounded memory (say, all memory-3 Markov
sources) as a natural hierarchy of classes. As the prior example
shows, these classes are definitely not natural from an estimation
point of view. Small memory sources---even with memory one can be
arbitrarily slow mixing---and hence hard to estimate. On the other
hand, sources with longer memory may be easier to estimate if they are
fast mixing and satisfy certain other conditions, as we will see
below. 

As a consequence, we look for a different way to resolve the class of
all finite memory Markov sources into a nested hierarchy of classes.
Therefore, in~\cite{asadi2014stationary}, a new class of Markov sources
was introduced, one that was more amenable to estimation. These
classes of sources were not bounded in memory, rather they can have
arbitrarily long memory.  However, these sources satisfy a continuity
condition~\cite{csiszar2006context,asadi2014stationary}, as described
technically in Section~\ref{s:cnt:cont} in the paper.  

Roughly speaking, the continuity condition imposes two intuitive
constraints closely related to each other.
Let $p$ be a binary Markov source with finite but unknown
memory, and consider $p(X_0|X^{-1}_{-\infty})$. Because the source has
finite memory, there is a suffix of the past, $X_{-\infty}^{-1}$ that
determines the conditional probability above. Since we do not
have an a-priori bound on the memory, we cannot say how much of the
past we need. Yet, the conditional probabilities $p(X_0|X^{-1}_{-m})$
are well defined for all $m$. It is now possible to construct Markov
sources where, unless we have suffixes long enough to encapsulate the
true state (or equivalently, $m$ larger than the true memory),
$p(X_0|X^{-1}_{-m})$ is simply not a reflection of the true probability
$p(X_0|X^{-1}_{-\infty})$. 

The continuity condition prohibits this pathological
property---imposing on the other hand that the more of the context
($X^{-1}_{-m}$) we see, the better $p(X_0|X^{-1}_{-m})$ reflects
$p(X_0|X^{-1}_{-\infty})$.  Second, given a history $X^{-1}_{-m}$, the
continuity condition implies that the conditional information one more
bit in the past, $X_{-m-1}^{-1}$, provides on $X_0$ diminishes with $m$.

The continuity condition may be imagined as a soft constraint on the
memory, but it does not control mixing. Suppose we consider the
collection of all Markov sources that satisfy a given continuity
condition. It was shown in~\cite{asadi2014stationary} that it is possible to use
length-$n$ samples to estimate the conditional probabilities $p(1|\w)$
of all strings $\w$ of length $\log n$ that appear in the sample, as
well as provide deviation bounds on the estimates. 

This hints that Markov sources nested by the continuity condition (as
opposed to memory) are a natural way to break down the collection of
all Markov sources. In order to better understand these model classes,
in this paper we study compressing Markov sources constrained by the
continuity condition, and obtain asymptotically tight bounds on their
redundancy. 

Part of the reason to study this is to understand what portions of the
model classes are more important (namely, contribute primarily) to the
redundancy. Indeed, it turns out that the primary contribution to the
redundancy comes from essentially fast mixing sources whose state
probabilities that are not towards the extremes (not near 0 or 1), 
while, of course, these sources do not complicate estimation at all.
On the other hand, slow mixing sources hit estimation, but do not 
matter for compression at all---a dichotomy which was hinted at with
our very first memory-1 example. 

Our main results are matching lower and upper bounds, in Sections~\ref{s:lb}
and~\ref{s:ub} respectively. In Section~\ref{s:setup}, we set up notations,
and briefly describe the continuity condition in~\ref{s:cnt:cont}.

\subsection{Prior Work}
Davisson formulated the average and worst case redundancy in his
seminal paper in 1973~\cite{dav73}.  A long sequence of work has
characterized the worst case redundancy for memoryless sources
~\cite{kt81}~\cite{STW95}~\cite{xb97}~\cite{szp98}.  Average
redundancy of Markov sources with fixed memory has been studied
in~\cite{trofimov197}~\cite{davisson1983minimax}~\cite{atteson1999asymptotic}.
In~\cite{ris96}, the authors obtain the worst case redundancy of such
Markov sources and later~\cite{jacquet2004markov} derived the exact
worst case redundancy of such Markov sources. The estimation and
compression of finite memory context tree models was studied
in~\cite{willems1995context} and~\cite{rissanen1983universal}. \cite{csiszar2006context} and \cite{willems1998context} studied the estimation of context
tree with unbounded memory.

For a different comparison of estimation and compression in Markov
settings, see~\cite{falahatgar2016learning}. Here the authors obtain
the redundancy of conditionally describing one additional symbol after
obtaining a sample.  Finally,~\cite{dhulipala2006universal} considers
compression of patterns of Hidden Markov models.

\newcommand{\X}{{\mathbf X}}
\section{Setup and Notation}
\label{s:setup}
We denote length-$n$ strings in bold face $\x$ or as $x_1^n$, 
and their random counterparts are $\X$ and $X_1^n$ respectively.
For any $\x\in\{0,1\}^n$ and $\w\in\{0,1\}^{\ell}$ with $\ell\le n$,
let $n_{\w}$ be the number of appearances of $\w$ in $\x$ as a
\emph{substring}. For $\x,\y\in \{0,1\}^*$, $\x\prec \y$ denotes that
$\x$ is a suffix of $\y$ (\eg $010\prec 11010$),
$\x\y$ is the concatenation of $\x$ followed by $\y$ (\eg if $\x=010$
and $\y=1$ then $\x\y=\x1=0101$), and $|\x|$ as the length of $\x$.

We consider a two-sided infinite binary Markov random process $p$
generating a sample $\cdots,X_{-1}, X_0,X_1,\cdots$. The memory of $p$
is finite (though not bounded a-priori), \ignore{namely there exists
  some $\ell\in\mathbb{N}$ and for all $t\in\mathbb{Z}$, we have
$$p(X_{t}\mid X^{t-1}_{-\infty})=p(X_{t}\mid X_{t-1}^{t-\ell}),$$we say the minimal $\ell$ is memory of $p$.} and let $S_p$ to be \emph{context tree}~\cite{STW95} of source $p$. $S_p$ is a set of leaves of a complete binary tree and $p$ is completely described by the conditional (transition) probabilities $p(1\mid \s)$ for $\s\in S_p$. 

Let $\M^{\ell}$ to be all the Markov chains with memory at most
$\ell$, and $\M=\cup_{\ell} \M^{\ell}$ be the family of all the finite
memory Markov chains. As mentioned before, while $\M^\ell$ is a
natural class, we are looking to break $\M$ down into a more natural
hierarchy of classes.

\subsection{Markov chain with continuity condition}
\label{s:cnt:cont}
Let $\delta:=\mathbb{N}\rightarrow \mathbb{R}^+$ be a function such
that $\delta(n)\rightarrow 0$ as $n\rightarrow \infty$. A Markov
chain $p$ satisfies the \emph{continuity condition} subject to
$\delta$ if for all $\s_1,\s_2\in S_p$, and $a\in\{0,1\}$, we have
\[
\left|\frac{p(a\mid \s_1)}{p(a\mid \s_2)}-1\right|\le \delta(|\w|)
\]
for all $\w\in \{0,1\}^*$ such that $\w\prec\s_1$ and $\w\prec\s_2$
(namely $\w$ is a common suffix of $\s_1$ and $\s_2$). For technical reasons, we will assume that $\delta(n)<\frac{1}{4n}$, see \cite{asadi2014stationary}.

Denote $\M_{\delta}$ to be the family of all the Markov chains with
continuity condition subject to $\delta$ and
$\M_{\delta}^{\ell}=\M_{\delta}\cap \M^{\ell}$. Roughly speaking, the
continuity condition constraints the transition probabilities of
states with long common suffix to be close. \ignore{The influence of past
to the next bits will dismissed according to $\delta$.} 

Let $S_p(\w)=\{\s\in S_p: \w\prec \s\}$. Clearly we have that the
stationary probability $p(\w)= \sum_{s\in S_p(\w)}p(\s)$ and that
$p(1|\w)= \Paren{\sum_{s\in S_p(\w)} p(1|\s) p(\s)}/p(\w)$. 
Let $p$ have memory $\ell$. In analogy to the true conditional probability
$p(1|\w)$, for a given $\x\in \{0,1\}^{n}$ and
past $x_{-\ell+1}^{0}$, let
\begin{equation}
\label{ptilde}
\tilde{p}(1\mid \w)
=
\frac{\displaystyle\sum_{s\in S_p(\w)}n_{\s}p(1\mid \s)}{n_{\w}},
\end{equation}
be the \emph{empirical aggregated distribution} of $p$, write it as $\tilde{p}_{\w}$ for simplicity. In a slight
abuse of notation here, $n_\s$ (respectively $n_\w$) is the number of
bits in $\x$ with context $\s$ (respectively $\w$), \ie number of bits
in $\x$ immediately preceded by $\s$ (respectively $\w$) when taking
the past $x_{-\ell+1}^0$ into account).  \ignore{ in terms of
  $\x_{-\ell}^{n}$.  We often write it is as $\tilde{p}_{\w}$ for
  simplicity if the $p$ and $\x_{-\ell}^n$ is clear in the context.}

\subsection{The redundancy}
For any distribution family $\mathcal{P}$ on $\{0,1\}^n$, the worst case minimax redundancy of $\mathcal{P}$ is defined as
$$R(\mathcal{P})=\inf_{q}\sup_{p\in\mathcal{P}}\max_{\x\in\{0,1\}^n}\log\frac{p(\x)}{q(\x)},$$similarly, the average minimax redundancy is defined as
$$\tilde{R}(\mathcal{P})=\inf_q\sup_{p\in \mathcal{P}}E_{\X\sim p}\log\frac{p(\X)}{q(\X)},$$where $q$ is choosing from all the possible distributions on $\{0,1\}^n$. For any given (\emph{fixed}) past $x_{-\infty}^0$, we know that for any $p\in \M$ we will have a well defined distribution over $\{0,1\}^n$, given by
\[
\bar{p}(x_1^n)=p(x_1^n\mid x_{-\infty}^0).
\] 
\ignore{Since the past is \emph{fixed}, we know that $\bar{p}$ is unique determined by $p$, in the following context we will use $p$ as $\bar{p}$ without loss of generality.} 
The main result of this paper is a lower and upper bound on the average redundancy of $\M_{\delta}$ over $\{0,1\}^n$ for any  past, i.e.
$$\tilde{R}(\M_{\delta})\overset{\Delta}{=}\inf_q\sup_{p\in \M_{\delta}}\sup_{x_{-\infty}^0}E_{\X\sim \bar{p}}\log\frac{\bar{p}(\X)}{q(\X)}.$$
\section{The lower bound}
\label{s:lb}
The \emph{Redundancy-Capacity} theorem~\cite{csiszar2004information}
is a common approach to lower bound the minimax redundancy. This approach gets complicated in our case since there is no
universal bound on the memory of sources in $\M_{\delta}$, rendering
the parameter space to be infinite dimensional. We
therefore first consider $\M^{\ell}_\delta$ (see Section II), the
subset of sources in $\M_\delta$ which also have finite memory
$\le \ell$, and obtain a lower bound on
$\tilde{R}(\M^{\ell}_{\delta})$. Since
$\tilde{R}(\M_{\delta})\ge \tilde{R}(\M^{\ell}_{\delta})$ for all
$\ell$, we optimize over $\ell$ to obtain the best possible lower
bound on $\tilde{R}(\M^{\ell}_{\delta})$. Recall that
$\delta(\ell)\le 1/(4\ell)$.

\bTheorem[Lower Bound]\label{thm:lb} For any $\ell$, we have
\[
\tilde{R}(\M_{\delta}^{\ell})
\ge 
2^{\ell-1}\log n-2^\ell(\log \frac1{\delta(\ell)}+ \ell/2)-2^{\ell-1}(\log 4\pi e\ell+1),
\]
and $\tilde{R}(\M_{\delta})\ge \max_{\ell}\tilde{R}(\M_{\delta}^{\ell})$.
\ignore{Thus
\begin{align*}
\tilde{R}(\M_{\delta})&\ge \max_{\ell}\tilde{R}(\M_{\delta}^{\ell})\\
&=\max_{\ell}2^{\ell-1}\log n
-2^{\ell}\left(\log\frac{1}{\delta(\ell)}+(1+o_\ell(1))\ell/2\right).
\end{align*}}
\eTheorem

Before proving this theorem, we consider specific forms for $\delta$
to get an idea of the order of magnitude of the redundancy in Theorem~\ref{thm:lb}.
\bCorollary
If $\delta(\ell)=\frac{1}{\ell^c}$ with $c>1$, then we have
\[
\tilde{R}(\M_{\delta})=\Omega(n/\log^{2c-1}n),
\]
for $\ell=\log n-2c\log\log n+o(1)$.
If $\delta(\ell)=2^{-c\ell}$, then
\[
\tilde{R}(\M_{\delta})=\Omega(n^{1/(2c+1)}\log n),
\]
for $\ell=\frac{1}{2c+1}\log n+o(1)$. 
If $\delta(\ell)=2^{-2^{c\ell}}$, then
\[
\tilde{R}(\M_{\delta})=\Omega(\log^{1+1/c}n),
\]for $\ell = \frac{1}{c}\log \log n+o(1)$.
\eCorollary
For any $p\in \M_{\delta}^{\ell}$, we know that the distribution of $p$ on $\{0,1\}^n$ can be uniquely determined by at most $2^{\ell}$ parameters, i.e. the transition probabilities $p(1\mid \s)$. Let
$$\hat{\M}_{\delta}^{\ell}=\{p\in \M_{\delta}^{\ell}\mid \forall \s\in S_p, ~|p(1\mid \s)-1/2|\le \delta(\ell)\},$$be a sub-family of $\M_{\delta}^{\ell}$ with all the transition probabilities close to $1/2$. The following lemma is directly from \emph{Redundancy-Capacity} theorem\cite{csiszar2004information}.
\bLemma
\label{firstLem}
Let $\epsilon_\s$ be the maximum mean square error of estimating parameters $p\in\hat{\M}^{\ell}_{\delta}$ from their length $n$ sample. Then
\begin{equation}
\tilde{R}(\hat{\M}^{\ell}_{\delta})\ge2^{\ell}\log \delta(\ell) - 2^{\ell-1}\log \left(2\pi e\epsilon_\s\right).
\end{equation}
\Proof

Consider the following Markov chain
\[
\hat{\M}_{\delta}^{\ell}\overset{(a)}{\rightarrow} P \overset{(b)}{\rightarrow} X_{-\ell+1}^n\overset{(c)}{\rightarrow} \hatp,
\]
where $(a)$ $P$ is a random Markov process chosen from a uniform prior
over $\hat{\M}_{\delta}^{\ell}$, $(b)$ $X_1^n$ is a length $n$ sample
from distribution $P$, $(c)$ $\hatp$ is an estimate of $P$ from the
sample $X_1^n$ that uses the empirical probabilities
$\frac{n_{\s1}}{n_{\s}}$ to estimate $P(1\mid \s)$ for any
$\s\in\sets{0,1}^\ell$. 

 By capacity-redundancy theorem one knows that
$$\tilde{R}(\hat{\M}_{\delta}^{\ell})\ge I(P;X_1^n)\ge I(P;\hatp),$$where the second inequality is by data processing inequality. Note that
\begin{align}
\label{eqCapRed}
I(P;\hatp)&=h(P)-h(P|\hatp)\nonumber\\
&=h(P)-h(P-\hatp|\hatp)\nonumber\\
&\ge h(P)-h(P-\hatp),
\end{align}
where the last inequality follows since conditioning reduces entropy. 
To bound first term in~\eqref{eqCapRed} let $P\in\hat{\M}_{\delta}^{\ell}$ be uniform on the hypercube $A$ with edge lengths $\delta(l)$. Then
\label{enterlower}
\[
h(P)= 2^\ell\log\delta(\ell).
\]
since $h(P)=\log \text{Vol}(A)$.

To bound $h(P-\hatp)$, let $K$ be the covariance matrix of any estimator of parameter space condition on $\x^n$. Then using Theorem 8.6.5 in~\cite{CT}
\[
h(P-\hatp)\leq \frac1{2}\log (2\pi e)^{2^\ell}|K|.
\]
Let $|K|$ and $\lambda_i$ show determinant and eigenvalues of matrix $K$, respectively. Let $\epsilon_i$ be the element diagonal elements of covariance matrix. Then by the definition of trace of a matrix 
\[
\sum_i\epsilon_i= \mathrm{tr}({K})=\sum_i {\lambda_i}.
\]
Using arithmetic-geometric inequality, we get
\begin{equation}
\big(\frac{\sum_i{\lambda_i}}{2^\ell}\big)^{2^\ell}\geq \prod_i {\lambda_i}.
\end{equation}
Also $\sum_i\epsilon_i\leq 2^\ell\epsilon_\s$. Then
\begin{equation}
\label{eql1}
|K|=\prod_i {\lambda_i}\leq \big(\frac{\sum_i\epsilon_i}{2^\ell}\big)^{2^\ell}\leq \epsilon_\s^{2^\ell}.
\end{equation}
Applying~\eqref{eql1} in $\frac1{2}\log (2\pi e)^{2^\ell}|K|$, we have
\[
h(P-\hatp)\leq 2^{\ell-1}\log \big(2\pi e \epsilon_\s\big).
\] 
and lemma follows. 
\eLemma

To bound $\epsilon_\s$ one needs to find an estimator that makes it as small as possible. We will show that the empirical estimation $\hatp(\s)=\frac{n_{\s1}}{n_{\s}},$ is sufficient to establish our lower bound. We now find an upper bound on the estimation error of state $\s$ using naive estimator. 
\bLemma 
\label{lerror}
 Consider the naive estimator $\hatp(\s)=\frac{n_{\s1}}{n_{\s}}$. Then, 
\[
E[(\hatp(\s)-P(\s))^2] \leq \min\{E\left[\frac{1}{n_{\s}}\right],1\}.
\]
\Proof
Note that
\[
E[(\hatp(\s)-P(\s))^2]  = E\big[ E[(\hatp(\s)-P(\s))^2|n_\s]\big].
\]
Condition on $n_\s$, the symbols follow string $\s$ can be considered as outputs of an \iid Bernoulli with parameter $P(\s)$. For a sequence of zeros and ones with length $n$ drawn \iid from $B(p)$ with $k$ one, it is easy to see that $E[(\frac{k}{n}-p)^2]\leq \frac1{n}$, so
\[
E\big[ E[(\hatp(\s)-P(\s))^2|n_\s]\leq 
\min\{E\left[\frac{1}{n_{\s}}\right],1\}.
\]
\eLemma 
So finding the lower bound on redundancy reduces to find an upper bound on $E[\frac1{n_\s}]$. We need two following technical lemmas to bound $E[\frac1{n_\s}]$.
\bLemma
\label{techlemma}
Let $X_1,X_2,\cdots,X_n$ be binary random variables, such that for any $1\le t\le n$
$$\mathrm{Pr}\left(X_t=1\mid X_1,\cdots,X_{t-1}\right)\ge q,$$for some $q\in [0,1]$. Then, for any $1\le k\le n$
$$\mathrm{Pr}\left(\sum_{i=1}^nX_i\le k\right)\le \sum_{i=0}^{k}\binom{n}{i}q^i(1-q)^{n-i}.$$
\Proof
We use double induction on $k$ and $n$ to prove this theorem. 

Consider the base case for $k=0$, and an arbitrary $n$, then we need to bound $\mathrm{Pr}\left(\sum_{i=1}^nX_i\le 0\right)$, which is equal to say that  $\Pr(X_1=0,X_2=0,\dots,X_n=0)$. But
\begin{align*}
\Pr(X_1=0,\dots,X_n=0)
&=\Pr(X_n=0|X_1=0,\dots,X_{n-1}=0)\dots\Pr(X_1=0)\\
&\leq(1-q)^n
\end{align*}
where the first equation is using chain rule and the inequality follows by the assumption that 
\[
\mathrm{Pr}\left(X_n=1\mid X_1,\cdots,X_{n-1}\right)\ge q.
\]

If $k= n$ then
\[
\sum_{i=0}^{n}\binom{n}{i}q^i(1-q)^{n-i}=1
\]
so we need to have $\mathrm{Pr}\left(\sum_{i=1}^nX_i\le k\right)<1$ which holds always.

For the induction step we just need to show that if $(n',k')\leq (n,k)$ and
$$\mathrm{Pr}\left(\sum_{i=1}^{n'}X_i\le k'\right)\le \sum_{i=0}^{k'}\binom{n'}{i}q^i(1-q)^{n'-i},$$ 
holds, then 
 $$\mathrm{Pr}\left(\sum_{i=1}^nX_i\le k\right)\le \sum_{i=0}^{k}\binom{n}{i}q^i(1-q)^{n-i}$$
 holds. 
To see it, define 
$$A_k^n= \{\sum_{i=1}^nX_i\leq k\},$$
$$B_k^n=\{X_1=0 \wedge \sum_{i=1}^nX_i\leq k\},\text{\quad and}$$
$$C_k^n=\{X_1=1\wedge \sum_{i=1}^nX_i\leq k-1\}.$$
Define 
\[
T_k^n=\sum_{i=0}^k \binom{n}{i}q^i(1-q)^{n-i}.
\]
Using chain rule we have 
\[
\Pr\{B_k^n\}=\Pr\{X_1=0\}\Pr\{\sum_{i=1}^nX_i\leq k|X_1=0\}\]
\[
\Pr\{C_k^n\}=\Pr\{X_1=1\}\Pr\{\sum_{i=1}^nX_i\leq k-1|X_1=1\}.
\]
Let $P(X_1=1)=\tilde{q}>q$, then $P(X_1=0)=1-\tilde{q}<1-q$. 

Note that $A_k^n=B_k^n\cup C_k^n$. Using union bound and since $B_k^n$ and $C_k^n$ and are disjoint, 
\[
\Pr\{A_k^n\}=\Pr\{B_k^n\}+\Pr\{C_k^n\},
\]
Then 
\begin{align*}
\Pr\{A_k^n\}
&=(1-\tilde{q})\Pr\{\sum_{i=1}^nX_i\leq k|X_1=0\}+\tilde{q}\Pr\{\sum_{i=1}^nX_i\leq k-1|X_1=1\}\\
&\overset{(a)}{\leq} (1-q)T_{k}^{n-1}+qT_{k-1}^{n-1}\\
&=(1-q)\sum_{i=0}^{k}\binom{n-1}{i}q^i(1-q)^{n-1-i}
+q\sum_{i=0}^{k-1}\binom{n}{i}q^i(1-q)^{n-i}\\
&=\sum_{i=0}^{k}\binom{n}{i}q^i(1-q)^{n-i}
\end{align*}
and $(a)$ follows by using induction assumption. 

\eLemma

\bLemma
\label{probLem}
For all $p\in \hat{\M}_{\delta}^{\ell}$, we have
\[
p\bigg(n_\s\leq\frac{n}{2\ell 2^\ell}-\sqrt{\frac{n\log n}{\ell 2^{\ell}}}\bigg)\leq \frac1{n},
\] 
\Proof 
Divide length $n$ sequence to subsequence of length $l$ and let $m=\frac{n}{\ell}$. Let
$1_i$ for $i\in\{1,2,\dots,m\}$ as
\[
1_{i}=1\{\s\prec X_{(i-1)\ell}^{i\ell}\}
\]
Note that 
\begin{align*}
p(1_{i}=1\mid 1_{0},1_{1},\cdots,1_{i-1})
&\overset{(a)}{\ge} \left(\frac{1}{2}-\delta(\ell)\right)^{\ell}\\
&\ge \frac{1}{2^{\ell}}(1-2\delta(\ell))^{\ell}\\
&\overset{(b)}{\ge} \frac{1}{2^{\ell}}(1-2\ell\delta(\ell))\\
&\ge \frac{1}{2^{\ell+1}},
\end{align*}
where $(a)$ follows since $p(1\mid \s')\in [1/2-\delta(\ell),1/2+\delta(\ell)]$ for all $\s'\in S_p$ and $(b)$ is by union bound.
Let $q=\frac1{2^{\ell+1}}$ in Lemma~\ref{techlemma}, then
\[
\Pr\big(n_s\leq k\big)\leq  \sum_{i=0}^k{m \choose i}(\frac1{2^{\ell+1}})^i(1-{\frac1{2^{\ell+1}}})^{m-i}.
\]
Right hand side in last inequality is the probability that sum of some  \iid random variables (we denote it by $S$) drawn from $\mathcal{B}(\frac1{2^{l+1}})$ with mean $\mu=\frac{m}{2^{\ell+1}}$ is less than k. Let $k=(1-\gamma)\mu$ where $0 \leq \gamma\leq 1$ is arbitrary. Then
\[
\sum_{i=0}^k{m \choose i}(\frac1{2^{\ell+1}})^i(1-\frac1{2^{\ell+1}})^{m-i}=\Pr(S\leq (1-\gamma)\mu).
\]
Using Chernoff bound, we get
\[
\Pr(S\leq (1-\gamma)\mu)\leq e^{-\frac{\gamma^2\mu}{2}}.
\]
Let $\gamma=2\sqrt{\frac{2^{\ell}\ell\log n}{n}}$, then 
\[
 e^{-\frac{\gamma^2\mu}{2}}
 =e^{-2\frac{2^{\ell+1}\ell\log n}{n}\frac{n}{2^{\ell+1}\ell}}
 \leq \frac1{n}.
\]
So $k=(1-\gamma)\mu=\frac{n}{2^{\ell+1}\ell}-\sqrt{\frac{n\log n}{2^{\ell}\ell}}$ and lemma follows.
\eLemma

We now combine Lemma~\ref{techlemma} and Lemma~\ref{probLem} to bound $E\big[\frac1{n_\s}\big]$.
\bLemma For $n$ large enough,
\label{linvns}
\[
E\bigg[\frac1{n_\s}\bigg]\leq  \frac{\ell 2^{\ell+1}}{n}.
\]
\Proof
Let $k=\frac{n}{2^{\ell+1}\ell}-\sqrt{\frac{n\log n}{2^{\ell}\ell}}$, then
\begin{align*}
E\bigg[\frac1{n_\s}\bigg]
&=\sum_{\frac1{n_\s}\geq \frac1{k}}\frac1{n_\s}p\bigg(\frac1{n_\s}\geq \frac1{k}\bigg) 
	+\sum_{\frac1{n_\s}< \frac1{k}}\frac1{n_\s}p\bigg(\frac1{n_\s}< \frac1{k}\bigg)\\
&\leq \sum_{\frac1{n_\s}\geq \frac1{k}}p\bigg(\frac1{n_\s}\geq \frac1{k}\bigg)
	+\frac1{k}\sum_{\frac1{n_\s}< \frac1{k}}p\bigg(\frac1{n_\s}< \frac1{k}\bigg),
\end{align*}
where the inequality is by the fact that $\frac1{n_\s}<1$. Using Lemma~\ref{probLem}
\[
\sum_{\frac1{n_\s}\geq \frac1{k}}p\bigg(\frac1{n_\s}\geq \frac1{k}\bigg)<\frac1{n}.
\]
Also $\sum_{\frac1{n_\s}< \frac1{k}}p\bigg(\frac1{n_\s}< \frac1{k}\bigg)<1$. So	
\[
E\bigg[\frac1{n_\s}\bigg]
\leq
\frac1{n}+\frac1{k}.
\]
But $k=\frac{n}{2^{\ell+1}\ell}-\sqrt{\frac{n\log n}{2^{\ell}\ell}}=
\frac{n}{2^{\ell+1}\ell}(1-2\sqrt{\frac{2^{\ell}\ell\log n}{n}})
$, then 
\[
E\bigg[\frac1{n_\s}\bigg]
\leq 
\frac1{n}
+
\frac{\ell 2^{\ell+1}}{n}\bigg(\frac1{1-2\sqrt{\frac{2^{\ell}\ell\log n}{n}}}\bigg).
\]
But we can choose $n$ large enough so that $\sqrt{\frac{2^{\ell}\ell\log n}{n}}<\frac1{16}$, then
\[
\frac1{1-2\sqrt{\frac{2^{\ell}\ell\log n}{n}}}<2.
\]
So
\[
E\bigg[\frac1{n_\s}\bigg]\leq \frac{1+2\ell 2^{\ell+1}}{n}\simeq \frac{2\ell 2^{\ell+1}}{n}.
\]
\eLemma

We are now ready to give proof of theorem~\ref{thm:lb}.\\
\textbf{Proof of Theorem~\ref{thm:lb}}
\begin{align*}
\tilde{R}(\hat{\M}^{\ell}_{\delta})
&\overset{(a)}{\geq}
2^{\ell}\log \delta(\ell) - 2^{\ell-1}\log \left(2\pi e\epsilon_\s\right)\\
&\overset{(b)}{\geq} 
2^{\ell}\log \delta(\ell) - 2^{\ell-1}\log \big(2\pi e E[\frac1{n_s}] \big)\\
&\overset{(c)}{\geq}
2^{\ell}\log \delta(\ell) -2^{\ell-1}\log \big(2\pi e \frac{2\ell 2^{\ell+1}}{n} \big)\\
&= 2^{\ell-1}\log n -2^{\ell-1}\log 4\pi e\ell \\
&\quad\quad- 2^{\ell-1}\log 2^{\ell+1}-2^\ell\log \frac1{\delta(\ell)}\\
&=2^{\ell-1}\log n-2^\ell(\log \frac1{\delta(\ell)}+ \ell/2)-2^{\ell-1}(\log 4\pi e\ell+1)
\end{align*}
where $(a)$ is using Lemma~\ref{firstLem}, $(b)$ is by Lemma~\ref{lerror} and $(c)$ follows by Lemma~\ref{linvns}.

\section{The upper bound}
\label{s:ub}
To obtain the upper bound, we first show that for any given
$\x\in \{0,1\}^{n}$, the maximum probability from any distribution in
$\M_{\delta}$ will not much greater than that from $\M_{\delta}^{\ell}$ 
for an appropriate choice of
$\ell$.  This allows us a simple upper bound based on truncating
the memory of sources. Unfortunately (or fortunately), this simple
argument does not allow for tight matching bounds---hence we will need
to refine our argument further.

\bLemma \label{lemTruc}Fix any past $x_{-\infty}^0$. For any $\x\in\{0,1\}^n$, let
$\hat{p}(\x)=\max_{p\in \M_{\delta}}p(\x|x_{-\infty}^0)$ and
$\hat{p}_{\ell}(\x)=\max_{p\in \M_{\delta}^{\ell}}p(\x|x_{-\infty}^0)$. Then
$$\hat{p}(\x)\le 2^{2n\delta(\ell)}\hat{p}_{\ell}(\x).$$
\Proof
The continuity condition implies that for any $p\in \M_{\delta}$, we can find $p_{\ell}\in \M_{\delta}^{\ell}$ such that for all $a\in \{0,1\}$, $\w\in\{0,1\}^{\ell}$ and $\s\in S_p(\w)$, we have
\[
p(a\mid \s)\le (1+\delta(\ell))p_{\ell}(a\mid \w).
\] 
Thus we have $p(\x)\le (1+\delta(\ell))^np_{\ell}(\x)\le 2^{2n\delta(\ell)}\hat{p}_{\ell}(\x)$, where the last inequality follows since $(1+\delta(\ell))^{n}\approx e^{n\delta(\ell)}$.
\eLemma 
\bProposition 
$$R(\M_{\delta})\le \min_{\ell} 2^{\ell-1}\log n+2n\delta(\ell).$$
\Proof
Shtarkov's sum \cite{shtar1987universal} gives
\[
2^{R(\M_{\delta})}=\sum_{\x\in\{0,1\}^n}\hat{p}(\x).
\]
Thus by Lemma~\ref{lemTruc}, we have
\[
2^{R(\M_{\delta})}\le 2^{2n\delta(\ell)}\sum_{\x\in\{0,1\}^n}\hat{p}_{\ell}(\x)=2^{2n\delta(\ell)}2^{R(\M_{\delta}^{\ell})}.
\]
Therefore,
\[
R(\M_{\delta})\le 2n\delta(\ell)+R(\M_{\delta}^{\ell}).
\]
Observe that,
\[
R(\M_{\delta}^{\ell})\le R(\M^{\ell})\le2^{\ell-1}\log n,
\]
where the last inequality holds whenever $\ell>1$, see~\eg~\cite{STW95}.
\eProposition
As before, we work out the above bounds for specific $\delta$.
\bCorollary
If $\delta(\ell)=\frac{1}{\ell^c}$ with $c>1$, then
\[
R(\M_{\delta}) = O(n/\log^{c-1}n),
\]
for $\ell=\log n-c\log \log n+o(1)$.
For $\delta(\ell)=2^{-c\ell}$, we have
\[
R(\M_{\delta})=O(n^{1/(c+1)}\log n),
\]
for $\ell=\frac{1}{c+1}\log n+o(1)$.
For $\delta(\ell)=2^{-2^{c\ell}}$, we have
\[
R(\M_{\delta})=O(\log^{1+1/c} n),
\], for $\ell=\frac{1}{c}\log\log n$.
\eCorollary

Comparing Corollary 2 and Corollary 8, the upper and lower bounds on
the redundancy have asymptotically tight order when $\delta$  
diminishes doubly exponentially. 
For polynomial $\delta$ the lower bound and upper bound orders
differ by $\log n$ factors. 
However, for $\delta$ to be exponential, we have a
polynomial gap between the lower and upper bound. 

This suggests that either the lower or upper or both bound are too
loose. For the lower bound, recall that the main contribution came from
the fast mixing sources in $M$, while the other sources---the ones that
are problematic to estimate, were summarily ignored. 

Yet we will show in what follows that our lower bound given in Theorem
1 is actually tight. We need the following technical lemmas to refine 
our upper bound

\bLemma[Extended Azuma inequality]
\label{exAzuma}
Let $X_1,\cdots,X_k,\cdots$ be martingale differences with $|X_i|\le 1$, $\tau$ is a stopping time (i.e. event $\{\tau=k\}$ only depends on $\sigma(X_1,\cdots,X_k)$). If $\tau\le n$, then we have
$$\mathrm{Pr}\left(\left|\sum_{i=1}^{\tau}X_i\right|\ge \gamma \sqrt{\tau}\right)\le ne^{-\gamma^2/2}.$$
\Proof
Define $A_k=\{|X_1+\cdots+X_k|\ge \gamma \sqrt{k}\}$, $B_k=\{\tau=k\}$, let $C_k=A_k\cap B_k$. In fact, $C_n$ is the event that we stop at $n$ while it is a wrong time to stop. Note that $|X_i-X_{i-1}|<1$, using Azuma inequality we have
\[
\Pr[A_n]=\Pr\{|\sum_{i=1}^n X_i|\geq \gamma \sqrt{n}\}\leq e^{-\gamma^2/2}.
\]
Then 
\[
\Pr[\cup_{k=1}^n C_k]\leq \sum_{k=1}^n\Pr[A_k\cap B_k]\leq \sum_{k=1}^n \Pr[A_k]\leq n e^{-\gamma^2/2}.
\]
and the Lemma follows. 
\eLemma

\bLemma
\label{lemBoundEps}
For any $p\in \M_{\delta}$, we have
$$p\left(\sum_{\w\in\{0,1\}^{\ell}}|n_{\w1}-n_{\w}\tilde{p}_{\w}|\le \log n\sqrt{n2^{\ell}}\right)\ge 1-\frac{2^{\ell}}{n^3},$$for $n$ large enough that $\log n\ge 6$, where $\tilde{p}_{\w}$ is defined in Section II.
\Proof
Define 
\[
1_i(\s)=
\left\{ 
\begin{array}{c}
1,\quad \text{the $i$-th appearance of $\w$ in $\w\prec\s$ } \\ 
0, \quad  \text{otherwise.}
\end{array}
\right. 
\]
Let $W_i=\sum_{w\preceq s}1_i(\s)p(1|\s)$, and define
\[
Y_i(\w)=
\left\{ 
\begin{array}{c}
1, \quad \text{the $i$-th appearance of $\w$ happens follows by one} \\ 
0, \quad  \text{otherwise.}
\end{array}
\right. 
\]
Let $Z_i=Y_i-W_i$, then by Lemma 2 in~\cite{oshiro2017jackknife}, $Z_i$ are martingale differences and $|Z_i|<1$. 

Note that $n_{\w1}=\sum_i Y_i$ and $n_{\w}\tilde{p}_{\w}=\sum_i W_i$ and
\[
|n_{\w 1}-\tilde{p}_\w n_\w|=\sum_{\s\in S_\w(p)}|n_{\s 1}-p(1|\s)n_\s|=|\sum_{i=1}^{n_\w}Z_i|
\]
Define $z_\w=|n_{\w 1}-\tilde{p}_\w n_\w|$. Then using Lemma~\ref{exAzuma},
\begin{align*}
p\bigg( z_\w \geq \log n\sqrt {n_\w}\bigg)
&\leq ne^{-\log n^2/2}\\
&=ne^{\log n^{-\log n/2}}\\
&=\frac{n}{n^{\log n/2}}\\
&\leq \frac1{n^3}.
\end{align*} 

Let $A_\w=\{z_\w<\log n\sqrt{n_\w}\}$. Then 
\begin{align*}
p(\cup_\w A_\w^c)&=p (\cup_\w\{ z_\w\geq\log n\sqrt{n_\w}\})\\
&\leq\sum_{\w\in\{0,1\}^{\ell}} p(z_\w\geq\log n\sqrt{n_\w})\\
&\leq \sum_{\w\in\{0,1\}^{\ell}} \frac{1}{n^3}\\
&=\frac{2^\ell}{n^3}.
\end{align*}
Therefore, 
\[
p(\cap_{\w} A_\w)=1-p(\cup_\w A^c_\w)\geq 1-\frac{2^\ell}{n^3}.
\]
Note that event $\{\cap_{\w} A_\w\}$ implies event $\{\sum_{\w\in\{0,1\}^{\ell}} z_\w<\sum_{\w\in\{0,1\}^{\ell}} \log n\sqrt{n_\w}\}$, so 
\[
 p\bigg(\sum_{\w\in\{0,1\}^{\ell}} z_\w<\sum_{\w\in\{0,1\}^{\ell}} \log n\sqrt{n_\w}\bigg)\geq p (\cap_{\w} A_\w)\geq 1-\frac{2^\ell}{n^3}.
\]
Also, 
\begin{align*}
 p\bigg(\sum_{\w\in\{0,1\}^{\ell}} z_\w<\sum_{\w\in\{0,1\}^{\ell}} \log n\sqrt{n_\w}\bigg)
=p\bigg(\bigg(\sum_{\w\in\{0,1\}^{\ell}} z_\w\bigg)^2< \big(\sum_{\w\in\{0,1\}^{\ell}} \log n\sqrt{n_\w}\big)^2\bigg).
\end{align*}
Using Cauchy-Schwartz inequality we have,
\[
\bigg(\sum_{\w\in\{0,1\}^{\ell}} \log n\sqrt{n_\w}\bigg)^2\leq \sum_{\w\in\{0,1\}^{\ell}} \log^2n\sum_{\w\in\{0,1\}^{\ell}} n_\w=n2^\ell\log^2 n.
\]
So,
\[
 p\bigg(\sum_{\w\in\{0,1\}^{\ell}} z_\w<\sum_{\w\in\{0,1\}^{\ell}} \log n\sqrt{n_\w}\bigg)
 =  p\bigg(\sum_{\w\in\{0,1\}^{\ell}} z_\w< \sqrt{n2^\ell}\log n\bigg),
\]
and the Lemma follows. 
\eLemma

Consider a sample $\x$ from $p\in\M_\delta$, past $x^0_{-\infty}$ and
consider the empirical aggregated probabilities in~\eqref{ptilde} for
$\w\in\sets{0,1}^\ell$. We now consider a memory $\ell$ Markov source
that has its conditional probability of 1 given $\w\in\sets{0,1}^\ell$
equal to the empirically aggregated probabilities in~\eqref{ptilde},
call the source $\tilde{p}_\ell$.  Note that $\tilde{p}_\ell$ need not
be in the class $\M_\delta^\ell$ or $\M_\delta$, and while we do not
explicitly say so in notation for ease of readability,
$\tilde{p}_\ell$ depends on the sample $\x$.

For any $\x$ and $p\in\M_\delta$, and for $\w\in\{0,1\}^{\ell}$, let $z_{\ell}=\sum_{\w}z_{\w}$.

\bLemma
\label{lm18}
For any $p\in \M_{\delta}$ and $\x\in\{0,1\}^{n}$, we have 
$$\tilde{p}_{\ell+1}(\x)\le 2^{2n\delta^2(\ell)+2z_{\ell+1}\delta(\ell)}\tilde{p}_{\ell}(\x).$$
Moreover, we have
\[
p\left(\left\{\x:z_{\ell+1}\le 
\log n\sqrt{n2^{\ell+1}}\right\}\right)\ge 1-\frac{2^{\ell}}{n^3}
\]
\Proof 

Note that 
\[
\tilde{p}_{\ell+1}(\x)=\prod_{\w\in\{0,1\}^l}
\tilde{p}_{1\w}^{n_{1\w 1}}(1-\tilde{p}_{1\w})^{{n_{1\w}-n_{1\w 1}}}
\tilde{p}_{0\w}^{n_{0\w 1}}(1-\tilde{p}_{0\w})^{{n_{0\w}-n_{0\w 1}}}
\], and 
\[
\tilde{p}_{\ell}(\x)=\prod_{\w\in\{0,1\}^\ell}
\tilde{p}_{\w}^{n_{\w 1}}(1-\tilde{p}_{\w})^{n_{\w}-n_{\w 1}}
\]
So we just need to show that
\begin{align*}
\prod_{\w\in\{0,1\}^l}
\tilde{p}_{1\w}^{n_{1\w 1}}(1-)^{n_{1\w}-n_{1\w 1}}
\tilde{p}_{0\w}^{n_{0\w 1}}(1-\tilde{p}_{0\w})^{n_{0\w}-n_{0\w 1}}
\\
\leq 
2^{2n\delta^2(\ell)+2z_{\ell+1}\delta(\ell)}\prod_{\w\in\{0,1\}^\ell}
\tilde{p}_{\w}^{n_{\w 1}}(1-\tilde{p}_{\w})^{n_{\w}-n_{\w 1}}
\end{align*}
To see it, note 
\[
\tilde{p}_\w n_\w=\tilde{p}_{1\w}n_{1\w}+\tilde{p}_{0\w}n_{0\w}.
\]
Let 
\begin{align*}
\tilde{p}_{1\w}=\tilde{p}_\w+\tilde{p}_\w\delta_1,\\
\tilde{p}_{0\w}=\tilde{p}_\w+\tilde{p}_\w\delta_0,
\end{align*}
 for some $|\delta_1|<l$ and $|\delta_0|<l$. Then
\[
n_{0\w}\delta_0+n_{1\w}\delta_1=0.
\]
Let
\begin{align*}
n_{1\w 1}=\tilde{p}_{1\w}n_{1\w}+z'_1=\tilde{p}_\w n_{1w}+\tilde{p}_wn_{1w}\delta_1+z'_1,\\
n_{0\w 1}=\tilde{p}_{0\w}n_{0\w}+z'_0=\tilde{p}_\w n_{0\w}+\tilde{p}_\w n_{0\w}\delta_0+z'_0,
\end{align*}
for some $z'_0$ and $z'_1$. Also
\begin{align*}
 \log \tilde{p}_{1\w}^{n_{1\w 1}}(1-\tilde{p}_{1\w})^{n_{1\w}-n_{1\w 1}}
 &=
 n_{1\w 1}\log \tilde{p}_{\w}\\
 &\quad+ n_{1\w 0}\log (1-\tilde{p}_{1\w})\\
 &\leq
  n_{1\w 1} (\log \tilde{p}_{1\w}+\delta_1)\\
  &\quad+n_{1\w 0} (\log (1-\tilde{p}_\w)-\frac{\tilde{p}_\w}{1-\tilde{p}_\w}\delta_1))\\
  &=A_{1\w}+(\tilde{p}_\w+\frac{\tilde{p}_\w^2}{1-\tilde{p}_\w})n_{1\w}\delta_1^2\\
  &\quad+(\frac{\tilde{p}_\w}{1-\tilde{p}_\w})\delta_1z'_1\\
  &\leq A_{1\w}+2n_{1\w}\delta(\ell)^2+2\delta(\ell)|z'_1|.
\end{align*}
where  
\[
A_{1\w}=n_{1\w 1}\log\tilde{p}_{\w}+n_{1\w 0}\log(1-\tilde{p}_\w).
\]
Similarly,
\begin{align*}
 \log \tilde{p}_{0\w}^{n_{0\w 1}}(1-\tilde{p}_{0\w})^{n_{0\w}-n_{0\w 1}}
 &=
 n_{0\w 1}\log \tilde{p}_{0\w}+ n_{0\w 0}(1-\log \tilde{p}_{0\w})\\
 &\leq
  n_{0\w 1} (\log \tilde{p}_{\w}+\delta_0)\\
  &\quad+ n_{0\w 0} (\log  (1-\tilde{p}_\w)-\frac{\tilde{p}_\w}{1-\tilde{p}_\w}\delta_0))\\
  &=A_{0\w}+(\tilde{p}_\w+\frac{\tilde{p}_\w^2}{1-\tilde{p}_\w})n_{0w}\delta_0^2\\
  &\quad+(\frac{\tilde{p}_\w}{1-\tilde{p}_\w})\delta_0z'_0\\
  &\leq A_{0\w}+2n_{0\w}\delta(l)^2+2\delta(l)|z'_0|,
\end{align*}
and $A_{0\w}=n_{0\w 1}\log\tilde{p}_{\w}+n_{0\w 0}\log(1-\tilde{p}_\w)$. Summing over all $\w$, we have
\begin{align*}
&\sum_\w\log 
\tilde{p}_{1\w}^{n_{1\w 1}}(1-\tilde{p}_{1\w})^{n_{1\w}-n_{1\w 1}}
\tilde{p}_{0\w}^{n_{0\w 1}}(1-\tilde{p}_{0\w})^{n_{0\w}-n_{0\w 1}}
\\
&\leq\sum_\w (A_{1\w}+ A_{0\w})+2n\delta(\ell)^2+2\delta(\ell)z_{\ell+1}\\
&=\log \tilde{p}_{\w}^{n_{\w 1}}(1-\tilde{p}_{\w})^{n_{\w}-n_{\w 1}}
+2n\delta(\ell)^2+2\delta(\ell)z_{\ell+1},
\end{align*}
where we use the fact that $z_{\ell+1}=\sum_\w|z'_0|+|z'_1|$.
Also using Lemma~\ref{lemBoundEps} one can see that 
\[
p\left(\left\{\x:z_{\ell+1}\le 
\log n\sqrt{n2^{\ell+1}}\right\}\right)\ge 1-\frac{2^{\ell}}{n^3}
\]
\eLemma
\bLemma
For any $p\in \M_{\delta}$, we have
$$p\left(\left\{\x:p(\x)\le 2^{r_{\ell}}\tilde{p}_{\ell}(\x)\right\}\right)\ge 1-\frac{\sum_{k=\ell}^{2\ell}2^{k}}{n^3},$$
where
$$r_{\ell}=n\delta(2\ell)+\sum_{k=\ell}^{2\ell}2n\delta^2(k)+2\log n \sqrt{n2^{k+1}}\delta(k).$$
%
\Proof
Using Lemma~\ref{lemTruc} we have
\begin{align}
\label{eqTrunc}
p(\x)\leq \tilde{p}_{2\ell}(x)
2^{2n\delta(2\ell)}.
\end{align}
Also, 
\begin{align*}
\tilde{p}_{\ell+1}(\x)
&=\prod_{\w\in\{0,1\}^\ell}   
\tilde{p}_{1\w}^{n_{1\w 1}}(1-\tilde{p}_{1\w})^{n_{1\w}-n_{1\w 1}}
\tilde{p}_{0\w}^{n_{0\w 1}}(1-\tilde{p}_{0\w})^{n_{0\w}-n_{0\w 1}}
\\
&\leq 
2^{2n\delta^2(\ell)+2z_{\ell}\delta(\ell)}\prod_{\w\in\{0,1\}^{\ell}}
\tilde{p}_{\w}^{n_{\w 1}}(1-\tilde{p}_{\w})^{n_{\w}-n_{\w 1}}.
\end{align*}
Similarly,
\begin{align}
\label{eqres}
\tilde{p}_{2\ell+1}(\x)
&\leq 
\sum_{k=\ell}^{2\ell} 2^{2n\delta^2(k)+2z_{k+1}\delta(k)}\prod_{\w\in\{0,1\}^{k}}
\tilde{p}_{\w}^{n_{\w 1}}(1-\tilde{p}_{\w})^{n_{\w}-n_{\w 1}}
\nonumber\\
&= 2^{\sum_{k=\ell}^{2\ell}2n\delta^2(k)+2z_{k+1}\delta(k)}\tilde{p}_\ell(\x).
\end{align}
and using equation~\eqref{eqTrunc} and equation~\eqref{eqres}, we have
\[
p(\x)\leq \tilde{p}_\ell(\x)2^{n\delta(2\ell)+\sum_{k=\ell}^{2\ell}2n\delta^2(k)+2z_{k+1}\delta(k)}.
\]
but note that for all $k$
\[
p\left(\left\{\x:z_{k+1}\le 
\log n\sqrt{n2^{k+1}}\right\}\right)\ge 1-\frac{2^{k}}{n^3},
\]
Using union bound
\[
p\left(\left\{\x:\sum_{k=l}^{2l}z_{k+1}\le \sum_{k=l}^{2l}
\log n\sqrt{n2^{k+1}}\right\}\right)\ge 1-\sum_{k=l}^{2l}\frac{2^{k}}{n^3},
\]
and the lemma follows. 

\eLemma

\bTheorem[Improved Upper Bound]
Redundancy of $\mathcal{M}_\delta$ is upper bounded by
\[
\tilde{R}(\mathcal{M}_\delta)
\leq 
	2^{\ell-1}\log n
+n\delta(2\ell)+\sum_{k=\ell}^{2\ell}\big(n\delta^2(k)+\log n\sqrt{n2^k}\delta(k)\big)
	+(2^{2\ell+1}-2^\ell)\frac{n}{n^3}
\]
for any integer $\ell\in\naturals$.

\Proof
Let $\mathcal{T}_p=\{\x:p(\x)\leq 2^{r_\ell}\tilde{p}_\ell(\x)\}$ be the set of good sequences and $\mathcal{T}_p^c=\{\x:\x\notin \mathcal{T}\}$.
Let $c(\x)$ be the best code for memory $\ell$ sources. Let $|c(\x)|$ denote the length of $c(\x)$. Let $q(\x)=\frac{2^{-c(\x)}+2^{-n}}{2}$. We can choose $c(\x)$ tight enough so that $\sum 2^{-c(\x)}=1$.


%


Then

\begin{align*}
\tilde{R}(\M_{\delta})
&=\max_{p\in\mathcal{M}_\delta}
\sum_\x p(\x)\log\frac{p(\x)}{q(\x)}\\
&=\max_{p\in\mathcal{M}_\delta}\sum_{\x\in\mathcal{T}_p}p(\x)\log\frac{p(\x)}{q(\x)}
	+\sum_{\x\in\bar{\mathcal{T}_p}}p(\x)\log\frac{p(\x)}{q(\x)}\\
&\leq
\max_{p\in\mathcal{M}_\delta} \sum_{\x\in\mathcal{T}_p}p(\x)\max_{\x\in\{0,1\}^n}\log\frac{p(\x)}{q(\x)}
	+\sum_{\x\in\bar{\mathcal{T}}_p}p(\x)\max_{\x\in\{0,1\}^n}\log\frac{p(\x)}{q(\x)}\\	
&\leq \max_{p\in\mathcal{M}_\delta}
\sum_{\x\in\mathcal{T}_p}p(\x)\max_{\x\in\{0,1\}^n}\log\frac{p_{\ell}2^{r_\ell}}{q(\x)}
	+\sum_{\x\in\bar{\mathcal{T}}_p}p(\x).n\\
&\leq \max_{p\in\mathcal{M}_\delta} \max_{\x\in\{0,1\}^n}	\log\frac{p_{\ell}2^{r_\ell}}{q(\x)}
	+n\frac{\sum_{k=\ell}^{2\ell}2^k}{n^3}\\
&= \max_{p\in\mathcal{M}_\delta} \max_{\x\in\{0,1\}^n}	[\log p_{\ell}(\x)+c(\x)+1]+r_\ell
	+n\frac{\sum_{k=\ell}^{2\ell}2^k}{n^3}\\	
&=2^{\ell-1}\log n+{r_\ell}
+(2^{2\ell+1}-2^\ell)\frac{n}{n^3}.
\end{align*}
Where $r_\ell=n\delta(2\ell)+\sum_{k=\ell}^{2\ell}n\delta^2(k)+\log n	\sqrt{n2^\ell}\delta(k)$. Note that first term in the last equation follows since the worst case redundancy of Markov sources with memory $\ell$ and is bounded by $2^{\ell-1}\log n$. So
\begin{align*}
\tilde{R}(\mathcal{M}_\delta)
&\leq 
	2^{\ell-1}\log n
+n\delta(2\ell)\\
&\quad+\sum_{k=\ell}^{2\ell}\big(n\delta^2(k)+\log n\sqrt{n2^k}\delta(k)\big)
	+(2^{2\ell+1}-2^\ell)\frac{n}{n^3}.
\end{align*} 
\eTheorem
\bCorollary
For $\delta(\ell)=2^{-c\ell}$, we have
$$\tilde{R}(\M_{\delta})=O(n^{1/(2c+1)}\log n).\eqed$$

\Proof
Note that, 
\begin{align*}
\tilde{R}(\mathcal{M}_\delta)
&\leq 
	2^{\ell-1}\log n
+n\delta(2\ell)\\
&\quad+\sum_{k=\ell}^{2\ell}\big(n\delta^2(k)+\log n\sqrt{n2^k}\delta(k)\big)
	+(2^{2\ell+1}-2^\ell)\frac{n}{n^3},
\end{align*}
let $\delta(k)=2^{-ck}$ then
\begin{align*}
\sum_{k=\ell}^{2\ell}\delta^2(k)
&=\sum_{k=\ell}^{2\ell} 2^{-2ck}=2^{-2cl}+2^{-2c(l+1)}+\dots+2^{-2c(2\ell)}\\
&=2^{-2c\ell}\big(1+2^{-2c}+\dots+2^{-2c\ell}\big)\\
&=2^{-2c\ell}\frac{1-2^{-2c(\ell+1)}}{1-2^{-2c}}\\
&=\frac{2^{-2c\ell}-2^{-2c(2\ell+1)}}{1-2^{-2c}},
\end{align*}
and
\begin{align*}
\sum_{k=\ell}^{2\ell}2^{k/2}\delta(k)
&=\sum_{k=\ell}^{2\ell} 2^{(-c+1/2)k}\\
&=2^{(-c+1/2)\ell}+2^{(-c+1/2)(\ell+1)}+\dots+2^{(-c+1/2)2\ell}\\
&=2^{(-c+1/2)\ell}\big(1+2^{-c}+\dots+2^{-2c\ell}\big)\\
&=2^{(-c+1/2)\ell}\frac{1-2^{-c(\ell+1)}}{1-2^{-c}}\\
&=\frac{2^{(-c+1/2)\ell}-2^{(-2c+1/2)\ell-c}}{1-2^{-c}}.
\end{align*}
Let $\ell=c'\log n$, then
\begin{align*}
\tilde{R}(\mathcal{M}_\delta)
&\leq
\frac{n^{c'}}{2}\log n
+\frac{n}{n^{-2cc'}}\\
&\quad+n\bigg(\frac{n^{-2cc'}-\frac{n^{-4cc'}}{2^{-2c}}}{1-2^{-2c}}\bigg)\\
&\quad+\log n\sqrt{n}\bigg(\frac{n^{c'(-c+1/2)}-\frac{n^{(-2c+1/2)c'}}{2^c}}{1-2^{-c}}\bigg)\\
&\quad+(2n^{2c'}-n^{c'})\frac{n}{n^3}.
\end{align*}
Let $c'=\frac{1}{2c+1}$
\begin{align*}
\tilde{R}(\mathcal{M}_\delta)
&\leq
\frac1{2}n^{\frac{1}{2c+1}}\log n
+n^{\frac{1}{2c+1}}\\
&\quad+\frac{n^{\frac{1}{2c+1}}-\frac{n^{\frac{-2c}{2c+1}}}{2^{-2c}}}{1-2^{-2c}}\\
&\quad+\log n\bigg(\frac{n^{\frac{1}{2c+1}}-\frac1{2^c}n^{\frac{-c+1}{(2c+1)}}}{1-2^{-c}}\bigg)\\
&\quad+(2n^{\frac{2c}{2c+1}}-n^{\frac{1}{2c+1}})\frac{n}{n^3}\\
&=\mathcal{O}(n^{\frac{1}{2c+1}}\log n).
\end{align*}
\eCorollary
\section{Conclusion}
We proved matching (in order) upper and lower bounds on the redundancy
of universally compressing length-$n$ strings from $\M_\delta$. In the
process, we examined which sources contribute predominantly to the
redundancy---discovering that fast mixing sources, or where estimation
is uncomplicated by mixing biases, are the biggest contributors.  This
reveals an interesting dichotomy in the Markov setup---the sources
that make estimation complicated are different from the sources that
complicate compression.

In our future work, we hope to use these results to better understand
what sort of Bayesian priors work in the Markov setups for various
tasks, resampling techniques for Markov sampling---the initial results
of which are available in~\cite{oshiro2017jackknife}, as well as
extend our understanding of data-derived estimation~\cite{SA15:jmlr} into the
Markov regime.

\section*{Acknowledgment}
The authors were supported by NSF grants CCF 1619452 ``Slow mixing
Markov processes'' and by the Center for Science of Information
(CSoI), an NSF Science and Technology Center, under grant agreement
CCF-0939370.



\bibliographystyle{unsrt}
\bibliography{isit18}
\end{document}